# Simultaneous Localization of Electrons in different Δ-valleys in Ge/Si Quantum Dot Structures

Aigul Zinovieva[1, a *], Natalia Stepina[1,b], Anatoly Dvurechenskii[1,c], Leonid Kulik[2,d], Gregor Mussler[3,e], Juergen Moers[3,f], Detlev Grützmacher[3,g]

[1] Institute of Semiconductor Physics SB RAS, 630090 Novosibirsk, Russia

[2] Institute of Chemical Kinetics and Combustion SB RAS, 630090 Novosibirsk, Russia

[3] Peter Gruenberg Institute, Forschungszentrum Julich and Julich-Aachen Research Alliance

[a]aigul@isp.nsc.ru, [b]stepina@isp.nsc.ru, [c]dvurech@isp.nsc.ru, [d]chemphy@kinetics.nsc.ru, [e]G.Mussler@fz-juelich.de, [f]j.moers@fz-juelich.de, [g]d.gruetzmacher@fz-juelich.de.

**Abstract.** In the present work the possibility of simultaneous localization of two electrons in $\Delta^{100}$ and $\Delta^{001}$ valleys in ordered structures with Ge/Si(001) quantum dots (QD) was verified experimentally by the electron spin resonance (ESR) method. The ESR spectra obtained for the ordered ten-layered QD structure in the dark show the signal corresponding to electron localization in Si at the Ge QD base edges in $\Delta^{100}$, $\Delta^{010}$ valleys ($g_{zz}$=1.9985, $g_{in\text{-}plane}$=1.999). Light illumination causes the appearance of a new ESR line ($g_{zz}$=1.999) attributed to electrons in the $\Delta^{001}$ valley localized at QD apexes. The observed effect is explained by enhancement of electron confinement near the QD apex by Coloumb attraction to the photogenerated hole trapped in a Ge QD.

**Introduction**

One of the most promising systems for quantum logic operations is a system with semiconductor quantum dots (QD) [1]. Here the spin states of electrons localized in QDs can be considered as qubits. One-qubit operations can be carried out as the single spin rotations caused by short microwave pulses. For two-qubit operations, a sufficient tunnel coupling between neighboring qubits is required, then the electrons must be close enough to each other. To provide a selective access, these electrons should have different g-factors. The last requirement is long spin decoherence time. All these conditions can be satisfied by the QDs in a Ge/Si heterosystem. Ge QDs serve as stressors providing electron localization due to strain in the Si vicinity of Ge QDs [2]. Electron localization in Si can lead to long spin relaxation times due to a small spin-orbit interaction in this material and a small concentration of $^{29}$Si isotope with nonzero nuclear spin [3].

Electron localization in different regions near Ge/Si QDs, near the apexes and near the base edges of QDs provides different g-factors due to localization in different Δ-valleys [4]. In the latter work the electron localization of in different Δ-valleys was demonstrated for structures with double QD layers. It was established that the spatial localization of electrons, as well as the localization in the momentum space, depends on the spacer thickness between QD layers. In the structures with spacer thickness 2 nm electrons are localized at QD base edges, while localization at QD apexes is realized in the structures with 3 nm spacer. Simultaneous localization of electrons at the apexes and at QD base edges is still not observed. Such type of localization can provide a sufficient overlap of electron wave functions allowing the implementation of two-qubit operations based on the exchange interaction.

**Strain and g-factor of the electron localized in a Ge/Si heterosystem with QDs**

Strain in a Ge/Si(001) QD system can be used to obtain effective localization of electrons with different g-factors. Earlier D. K. Wilson and G. Feher demonstrated [5] the effect of the uniaxial strain on the g-factor value in ESR experiments for donors in Si. In this work the samples were subjected to external uniaxial stress (strain was of the order of $10^{-3}$) resulting in g-shift $g-g_0=10^{-4}$ due to the effect of valleys repopulation. Internal strain in a Ge/Si QD heterosystem is one order larger and allows one to obtain only two Δ-valley population (for example, $\Delta^{001}$ and $\Delta^{00-1}$ valleys) and the highest possible in this heterosystem g-factor difference $\delta g=1.1\cdot 10^{-3}$ (Ref.[6]). Such large g-factor anisotropy has been obtained in the recent ESR experiments [4] on the Ge/Si QD structures with localization of electrons near the QD apexes. The strain in this region is close to an effective uniaxial compression along the growth direction of the structure Z and an in-plane tension [7]. This strain causes splitting of the sixfold-degenerate Δ-valley and separation of two lower $\Delta^{001}$ and $\Delta^{00-1}$ valleys and four upper in-plane Δ-valleys. The localized electron state is formed by the states of two lower Δ-valleys. The symmetry of the g-tensor of this electron state is defined by the symmetry of isoenergetic surface (an ellipsoid of revolution). When external magnetic field **H** is applied parallel to the ellipsoid axis, the ESR signal with $g_\parallel=1.9995$ should be observed; when **H** is perpendicular to this axis, the ESR signal with $g_\perp=1.9984$ should be measured [4]. Another possible place of electron localization is the region close to the QD base edges [8]. At certain conditions the state of the electron localized in this region can be lower in energy scale than the state of the electron localized at the QD apex. The strain distribution in the Si region near the center of the QD base edge is close to the following: for the base edge directed along the [100] direction, there is a tension along the [100] direction and compression along the [010] direction. This leads to analogous splitting of the Δ-valley, only, in this case, the lower valleys will be $\Delta^{010}$ and $\Delta^{0-10}$ ones. In experiments this corresponds to the following g-factor values: $g_{zz}=1.9985$, $g_{in-plane}=1.999$.

The experimentally observed g-factor values can differ from the above-indicated above due to the penetration of the electron wave function in the region with a non-zero Ge content, as well as due to smallness of deformation-induced valley splitting. In the latter case the g-factor anisotropy can be less pronounced. However, if the relation $g_{zz} > g_{in-plane}$ is observed, it is possible to conclude that the electron is localized near the QD apex. In the opposite case, $g_{in-plane} > g_{zz}$, one can talk about electron localization near the QD base edges. So, the analysis of g-factor orientation dependence can be used to define the place of electron localization in the structures with Ge/Si QDs.

**Samples and experiment**

A sample with two-dimensional ordered arrays of Ge QDs was grown on Si(001) substrate by molecular-beam epitaxy and contains 10 QD layers. Each QD layer was formed during deposition of 4 Ge monolayers. The thickness of silicon layers between QD layers was 10 nm. The spacer layer was deposited in two stages: at the first stage, a 5 nm Si layer was deposited at temperature 370°C, at the second one the growth temperature was ramped from 370 to 525°C. This procedure was used to suppress the Ge-Si intermixing during QD overgrowth. The second and ninth spacer was Sb-doped with concentration $n\sim 10^{16}$ cm$^{-3}$. The structure was grown on substrate Si (100) with a resistivity of ~1000 Ohm·cm. On top of the structure, a 0.15 μm epitaxial n-Si layer (Sb concentration ~ $10^{16}$cm$^{-3}$) was grown. Scanning electron microscopy study of test structures with uncovered QDs on the surface showed that the QDs are close to squared pyramids with lateral size 35 nm and the QD positions correspond to pre-patterned pits with the distance between QD centers being 100 nm. Base edges of QDs are oriented along [100] and [010] directions.

ESR measurements were performed with a Bruker Elexsys 580 X-band EPR spectrometer using a dielectric cavity Bruker ER-4118 X-MD-5. The sample with QDs was cut along principal crystalline directions [110] and [-110] and glued on a quartz holder, allowing one to rotate the sample in the magnetic field. The entire cavity and the sample were maintained at low temperature (T=4.5K) with a helium flow cryostat (Oxford CF935). The angular dependencies of g-factor and ESR line width are obtained by rotating the magnetic field from [001] to [110] direction.

## Results and discussion

The ESR study was performed at different microwave power. At microwave power P=0.63μW (55db) the ESR data gives the richest information about the investigated sample. ESR spectra consist of two lines: the first line has $g_1$=1.9994, while the second one has g-factor $g_2$=1.9985 in the magnetic field **H ∥ Z**. Sample rotation leads to the broadening of the first line up to becoming undetectable, while the second line is narrowed and shifted toward lower magnetic fields. At the in-plane magnetic field applied parallel to the [110] direction, the g-factor of the second ESR line is $g_2$=1.999. The observed orientation dependence is typical for the electrons localized at the QD base edges (in $\Delta^{100}$ and $\Delta^{-100}$ valleys).

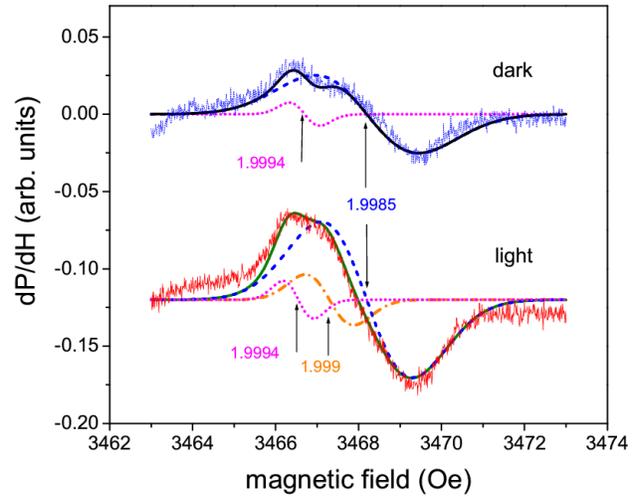

Fig. 1. ESR spectra of electrons localized in the structures with an ordered QD array with (without) illumination, T=4.5 K, microwave power P=0.63 μW, ν=9.70102 GHz, magnetic field is applied along the growth direction.

The orientation dependence of ESR line width $\Delta H_{pp}$ of the second line is very similar to that observed earlier for the QD structure with a large electron localization radius [9]. In the magnetic field parallel to the growth direction $\Delta H_{pp}$ was 2.1 Oe. With rotation of the sample in the magnetic field, the ESR line first narrows, and at $\theta$=30° the line width is reduced down to $\Delta H_{pp}$=1.4 Oe. Here $\theta$ is the angle between **H** and **Z**. With a further increase of angle $\theta$ the ESR line broadens and at $\theta$=60° the line width reaches $\Delta H_{pp}$=1.8 Oe. After $\theta$=60° the ESR line again narrows down to $\Delta H_{pp}$=1.1 Oe at $\theta$=90°. This orientation dependence can be explained by competition of two processes controlling the ESR line width. The first process results in the narrowing of ESR line due to the averaging of local magnetic fields produced by the $^{29}$Si nuclear spins. The second one leads to the broadening of ESR line due to a decrease of spin relaxation time $T_2$, $\Delta H_{pp}$~$1/T_2$. The decrease of $T_2$ with deviation of the external magnetic field from Z is typical for two-dimensional structures with structure-induced-asymmetry, where spin relaxation is controlled by Dyakonov-Perel (DP) mechanism. The important parameter that governs this competition is the effective electron localization radius $l$ (it is close to the size of QD base edge). This parameter in the structures under study is comparable with magnetic length $\lambda=(c\hbar/eH)^{1/2}$= 45 nm. When $l \sim \lambda$, the magnetic field applied along Z makes electron localization stronger through the effect of wave function shrinking. The deviation of the magnetic field from Z axis leads to an increase of wave function tails that provokes the electron hopping from one QD base edge to the other one. Such hops restricted within one QD lead to an effective averaging of the local magnetic fields and narrowing of the ESR line. At $\theta$>30° the shrinking effect vanishes and electrons can hop between QDs. This provides the broadening of ESR line due to the DP spin relaxation during hopping between QDs [9].

Fig. 1 demonstrates a change of ESR spectra under illumination in magnetic field **H ∥ Z**. The ESR spectrum in the dark consists of two lines with $g_1$=1.9994 (magenta dotted line) and $g_2$=1.9985 (blue dashed line). The first ESR line remains unchanged under illumination, then it is not related to QDs, while the intensity of the second ESR line doubles. Light illumination causes the appearance of a new ESR line ($g_3$=1.999, orange dash-dot line) attributed to the electrons in $\Delta^{001}$ ($\Delta^{00-1}$) valley localized at QD apexes. The intensity of this signal is much weaker than the intensity of ESR signal with g=1.9985 that can be interpreted as a consequence of a small potential well depth near the top of QD. Perhaps, the diffusive smearing of QD apex leads to a weakening of the confining potential for electrons in the $\Delta^{001}$ ($\Delta^{00-1}$) valley. The obtained g-factor value also confirms this assumption.

The nonzero Ge content in the electron localization area can provide g-factor value g=1.999. A simple estimation following expression $g_{el}=g(Si){\cdot}(1-\alpha) + g(Ge){\cdot}\alpha$, where $g(Si)$=1.9995, $g(Ge)$ = 1.9386 [9], $\alpha$ is Ge content in localization area, gives $\alpha$=0.8%, which is the reasonable value for our experimental structures. The weakness of ESR signal with g=1.999 does not allow us to study its orientation dependence, since this signal is broadened for other magnetic field orientations and cannot be resolved.

So, the obtained ESR data give the following information about the structures under study. The potential well at the QD base edge is deeper than the potential well at the QD apex. Thus, in the dark, electrons are mainly localized at the QD base edges in $\Delta^{100}$ ($\Delta^{-100}$), $\Delta^{010}$ ($\Delta^{0-10}$) valleys. Light illumination provides deepening of potential wells for electrons due to the Coloumb attraction to photogenerated holes trapped by Ge QDs. This enables simultaneous localization of electrons in different Δ-valleys at the base edges and the apexes of QDs.

## Summary

Simultaneous localization of electrons with different g-factors was observed in ESR study of ordered Ge/Si QD structures. Two ESR signals corresponding to the electrons localized at the apexes and at the base edges of QDs were detected under illumination. These electrons are located close to each other and can have a sufficient overlapping of wave functions. The last one can provide the exchange between them strong enough for organization of one- and two-qubit operations, and promote the quantum computing realization in the future.

This work is supported by RFBR (Grant 13-02-12105), SB RAS integration project No. 83 and DITCS RAS project No. 3.5 and Project ERA-NET-SB RAS (Grant 186).